\documentclass[]{article}

\pdfoutput=1
\usepackage{authblk}
\usepackage{fullpage}
\usepackage{amsmath}
\usepackage{booktabs}

\usepackage{caption}
\usepackage{graphicx}

\usepackage[colorlinks=True,
            citecolor=cyan,
            linkcolor=cyan,
            filecolor=cyan,
            urlcolor=cyan]{hyperref}
\usepackage[round, semicolon]{natbib}
\usepackage{multirow}
\usepackage{lscape}

\title{AutoHVSR: a machine-learning-supported algorithm
for the fully-automated processing of
horizontal-to-vertical spectral ratio measurements}
\author[1]{Joseph P. Vantassel}
\date{}
\author[2]{Andrew C. Stolte}
\author[2]{Liam M. Wotherspoon}
\author[3]{Brady R. Cox}
\affil[1]{Virginia Polytechnic Institute and State University}
\affil[2]{The University of Auckland}
\affil[3]{Utah State University}

\begin{document}

\maketitle

\begin{abstract}
The horizontal-to-vertical spectral ratio (HVSR) has seen widespread use as a tool for measuring a site's resonant behavior. However, the processing of HVSR with traditional methods can be tedious and time consuming when the HVSR is complex and exhibits multiple resonances. This work proposes the AutoHVSR algorithm that allows for fully-automated processing of HVSR measurements, including those with zero, one, or multiple clear resonances. The AutoHVSR algorithm accepts microtremor or earthquake recordings and user-defined HVSR processing parameters as input and returns the computed HVSR from each time window or earthquake record, the mean/median HVSR curve, and statistics on each automatically identified HVSR resonance in terms of frequency and amplitude. The AutoHVSR algorithm integrates robust signal processing and computational methods with state-of-the-art machine-learning models trained using a diverse dataset of 1109 HVSR measurements. The AutoHVSR algorithm demonstrates excellent performance by correctly determining the number of HVSR resonances for 1099 of the 1109 HVSR measurements (\textgreater 99\%) and predicting the mean resonant frequency of the correctly identified resonances with a root mean square error (RMSE) of 0.05 Hz. Furthermore, the AutoHVSR algorithm was able to produce these predictions in 13 minutes (including HVSR processing time) compared to the 30 hours required for traditional processing (a speed up of 138). The AutoHVSR algorithm is further demonstrated on a challenging dataset from Canterbury, New Zealand that included many HVSR curves with multiple and/or ambiguous resonances. Despite the challenging nature of the Canterbury dataset, the AutoHVSR algorithm was capable of correctly determining the number of HVSR resonances for 113 of the 129 HVSR measurements (\textgreater87\%) and predicting the mean resonant frequency of the correctly identified resonances with a RMSE of 0.10 Hz. The AutoHVSR algorithm produced these predictions under 2 minutes (including HVSR processing time) compared to the 4 hours required for traditional processing (a speed up of 120). Finally, while the AutoHVSR algorithm was developed using microtremor measurements where the horizontal components were combined using the geometric mean, it is shown to extend without modification to microtremor HVSR measurements where the two horizontal components are rotated azimuthally and to HVSR measurements from earthquake recordings. The AutoHVSR algorithm has been made publicly available in v0.3.0 of HVSRweb, it can be accessed at
\href{https://hvsrweb.designsafe-ci.org/}{https://hvsrweb.designsafe-ci.org/}.
\end{abstract}

\pagebreak

\section*{Introduction}

Since its conception \citep{nogoshi_amplitude_1971, nakamura_method_1989} the horizontal-to-vertical spectral ratio (HVSR) of microtremor data (i.e., ambient noise) has become an invaluable tool for seismic microzonation and site characterization \citep{ohmachi_ground_1991, konno_ground-motion_1998, fah_microzonation_1997, ibs-von_seht_microtremor_1999, oliveto_hvsr_2004, damico_ambient_2008}. The main advantage of HVSR when using microtremor recordings (i.e., mHVSR) over other non-invasive site characterization techniques is its ease of data acquisition. In particular, a single three-component seismometer, if left undisturbed and in good contact with the ground for times as short as 10-30 minutes (site's with lower resonant frequencies require longer recording times), can produce reliable estimates of a site's fundamental resonant frequency ($f_{0}$) (SESAME, 2004; Molnar et al., 2018, 2022). Some have even shown that HVSR measurements can be used to invert for a site's small-strain material properties, such as its shear wave velocity (Vs) profile \citep{fah_theoretical_2001, fah_inversion_2003, arai_s-wave_2004, herak_modelhvsrmatlab_2008, garcia-jerez_computer_2016, perton_inversion_2018}. Due to its acquisition simplicity and potential informative value, the utilization of mHVSR is expected to continue to grow, particularly through the acquisition of larger and more complex datasets aimed at characterizing spatial variability over large scales \citep{vantassel_mapping_2018, cheng_estimating_2021, hallal_hv_2021, panzera_reconstructing_2022, stolte_influence_2023} or across broad networks of ground motion recording stations \citep{wang_identification_2023}.

\begin{figure}[t!]
    \centering
	\includegraphics[width=1.0\textwidth]{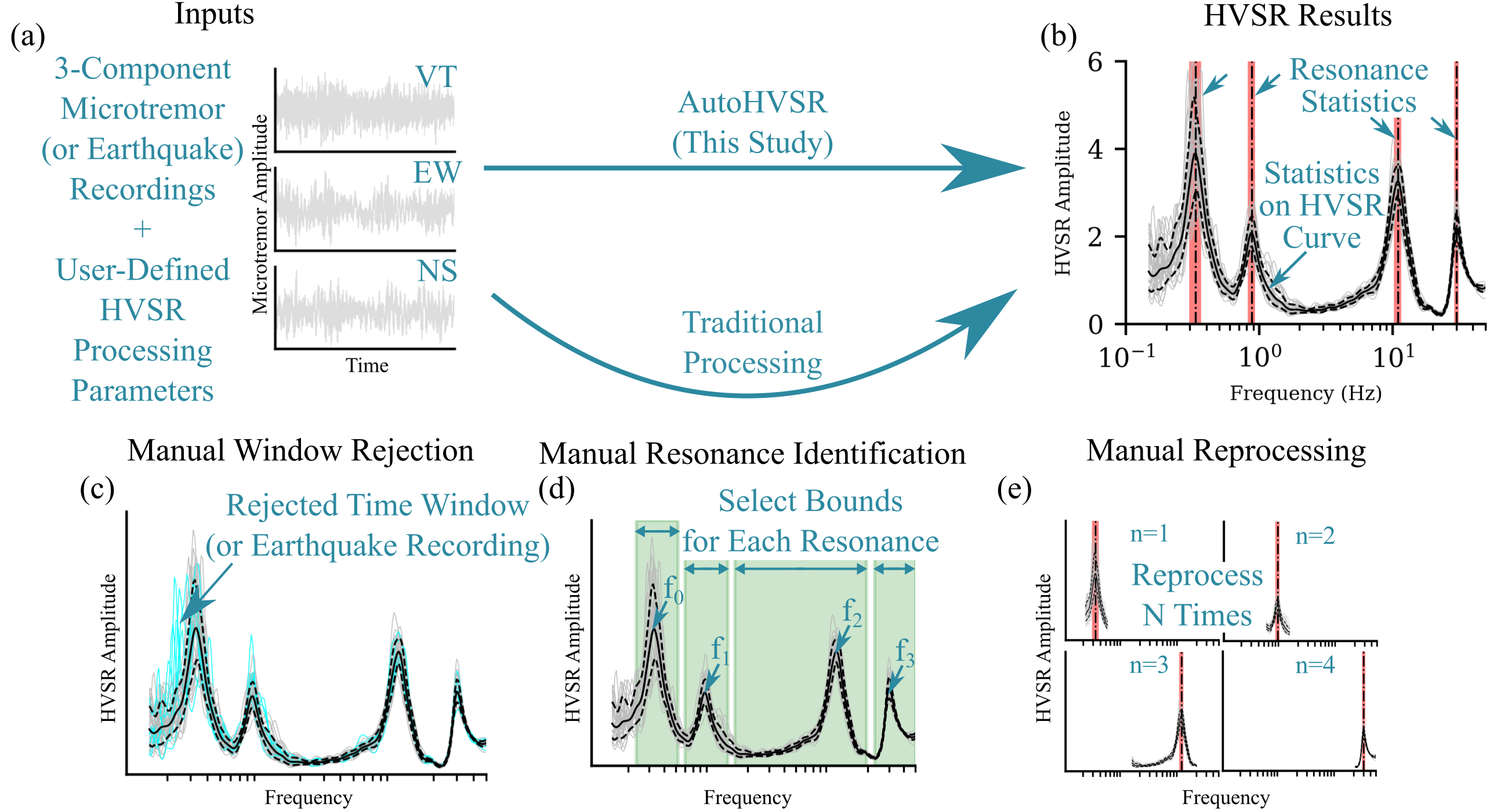}
	\caption{Comparison of traditional horizontal-to-vertical spectral ratio (HVSR) processing (bottom) and this study (top). While both approaches transform (a) 3-component microtremor (or earthquake; not shown) time-domain recordings to (b) HVSR results, traditional processing requires significant analyst interaction including (c) manual time window (or earthquake recording; not shown) rejection, (d) manual resonance identification, and (e) manual reprocessing. The present study, illustrated by the horizontal line at the top of the figure, seeks to replace the subjective and time-consuming analyst interaction of traditional processing with AutoHVSR, a machine-learning-supported algorithm for the fully-automated HVSR processing of microtremor and earthquake recordings.}
	\label{fig:1}
\end{figure}

The extraction of HVSR from microtremor data involves, first, dividing a three-component microtremor recording into a number of shorter time windows. These windows are typically between 30 and 120 seconds in length with longer windows being preferred at site's with lower $f_{0}$. Note that while the present discussion will focus on the processing of microtremor data, HVSR may also be calculated from earthquake recordings (i.e., eHVSR) \citep{lermo_site_1993, yamanaka_characteristics_1994, theodulidis_horizontal_1995, theodulidis_horizontal--vertical_1996, kawase_optimal_2011} with the only difference being that the aforementioned time windows in the mHVSR case are replaced by a set of earthquake recordings (i.e., one time window is analogous to one earthquake recording) in the eHVSR case. Therefore, when we refer to time windows throughout this work those interested in eHVSR can substitute the term “time window” for “earthquake recording” without loss of generality. After the division of a microtremor record, the time windows may then be de-trended, zero-padded, bandpass filtered, and/or cosine tapered. The horizontal components are then either transformed from the time domain to the frequency domain and combined, using for example the geometric mean, or combined in the time domain and then transformed to the frequency domain. After being transformed to the frequency domain, the vertical and horizontal spectra are smoothed to remove spurious peaks. The smoothed horizontal and vertical spectra from each time window are then divided to produce a set of HVSR curves (one per time window/earthquake recording). In traditional processing (e.g., SESAME \citeyear{sesame_guidelines_2004} ), the analyst must then visually review the set of HVSR curves, manually rejecting those that are not of good quality, and manually identify the apparent resonances. These steps, hereafter referred to as manual window rejection and manual resonance identification, respectively, are highly-subjective, time-consuming, and not easily scaled to large datasets. If one or more resonances are identified, the HVSR curves must be reprocessed, once per identified resonance, with an appropriately limited frequency range to isolate the resonance. Following reprocessing, statistics that represent the site's HVSR resonances in terms of frequency and amplitude are calculated from the peaks of the HVSR associated with each time window/earthquake recording. Finally, the HVSR curves are combined statistically, typically under the assumption of lognormality, to calculate a mean/median HVSR curve with an associated frequency-dependent standard deviation. An illustration of the traditional HVSR process with an HVSR curve with multiple clear peaks is shown in Figure \ref{fig:1}. In summary, HVSR processing accepts 3-component microtremor (or earthquake) recordings and user-defined processing parameters as input (Figure \ref{fig:1}a) and returns statistical representations of the HVSR curve and any apparent resonances (Figure \ref{fig:1}b). However, traditional processing requires significant analyst time for manual time window rejection (Figure \ref{fig:1}c), manual resonance identification (Figure \ref{fig:1}d), and manual reprocessing (Figure \ref{fig:1}e) making it difficult to perform quickly and reliably at scale.

To understand the importance of identifying all HVSR resonances, we present a brief overview of the disputed theoretical basis of the HVSR technique and its implications for automating HVSR processing. Nakamura in \citeyear{nakamura_method_1989} observed a correspondence between a site's horizontally-polarized vertically-propagating shear wave (SH-wave) transfer function and HVSR computed from microtremors. These observations led him to conclude that the HVSR curve is a direct measurements of a site's SH-transfer function \citep{nakamura_method_1989}, a position that subsequent publications confirm are still closely held \citep{nakamura_clear_2000, nakamura_what_2019} and shared by some \citep{herak_modelhvsrmatlab_2008} Later, other researchers began applying the HVSR technique to other real \citep{ohmachi_ground_1991, yamanaka_characteristics_1994, lermo_site_1994, fah_microzonation_1997, tokimatsu_geotechnical_1997} and simulated \citep{field_theoretical_1993, lachet_numerical_1994, dravinski_analysis_1996} microtremor datasets. Their findings indicated that while the frequency of $f_{0}$ agreed well between a site's theoretical SH-wave transfer function and HVSR from microtremors, the amplitude was not in good agreement \citep{field_theoretical_1993, lachet_numerical_1994, lermo_site_1994, dravinski_analysis_1996, tokimatsu_geotechnical_1997, fah_theoretical_2001, bonnefoy-claudet_hv_2006}. In addition, many noticed remarkable consistency between the peak in the fundamental mode Rayleigh wave ellipticity and the fundamental resonance of HVSR, leading to a new theory that the HVSR curve is primarily the result of Rayleigh wave ellipticity \citep{yamanaka_characteristics_1994, lachet_numerical_1994, fah_theoretical_2001, bonnefoy-claudet_hv_2006, castellaro_effect_2009}. This led to the first inversions of HVSR to estimate a site's Vs profile \citep{yamanaka_characteristics_1994, fah_theoretical_2001, fah_inversion_2003}. Shortly after, other studies began to consider both Rayleigh wave ellipticity as well as Love wave Airy phase as potential contributors to HVSR amplitude \citep{konno_ground-motion_1998, arai_s-wave_2004, bonnefoy-claudet_effects_2008}. The formative work by Konno and Ohmachi (\citeyear{konno_ground-motion_1998}) clearly demonstrated that for a site with a strong impedance contrast (\textgreater 2.5) a site's $f_{0}$ coincides with the peak of the fundamental mode Rayleigh wave ellipticity, the Love wave Airy phase, and the peak of the HVSR. More recently, a new theory based on the diffuse wavefield assumption \citep{sanchez-sesma_theory_2011, garcia-jerez_computer_2016, pina-flores_inversion_2017} has been proposed. In their original work, Sánchez-Sesma et al. (\citeyear{sanchez-sesma_theory_2011}) argue that the diffuse wavefield assumption \citep{sanchez-sesma_diffuse_2008} holds for microtremor data. If microtremor data can be modeled as a diffuse field, then the Green's function for the elastic media can be estimated from its cross-correlation. The Green's function can therefore be used as a bridge between experimental measurements of microtremors and the elastic material properties of the subsurface. Importantly, while the theoretical basis of HVSR may be disputed, all of the theories proposed to date can attribute meaning to some or all of the HVSR curve's amplitude. Furthermore, all of the theories broadly agree that a site with no clear resonance is representative of a relatively stiff subsurface (e.g., hard rock site) and one clear resonance is representative of one strong impedance contrast (e.g., soft soil over rock). However, they generally disagree about the informative potential of multiple resonance peaks. This is despite examples of multiple resonances in the HVSR curve appearing throughout the literature (even appearing in Nakamura's original work) but largely being downplayed or overlooked \citep{nakamura_method_1989, arai_s-wave_2004, bonnefoy-claudet_effects_2008, sanchez-sesma_diffuse_2008, sanchez-sesma_theory_2011}. Nevertheless, some have attempted to explain HVSR with multiple peaks in relation to multiple clear impedance contrasts at depth \citep{field_theoretical_1993, lachet_numerical_1994, bodin_microtremor_2001, asten_comment_2004, sesame_guidelines_2004, damico_ambient_2008, wotherspoon_site_2018, vantassel_mapping_2018, stolte_influence_2023}, however the informative potential of multiple resonance HVSR remains understudied. The authors expect that this is due in no small part to the difficulty of extracting multiple resonances from HVSR using current techniques.

Two previous semi-automated HVSR processing algorithms have been proposed in the literature, the first by Cox et al. (\citeyear{cox_statistical_2020}) and the second by Yazdi et al. (\citeyear{yazdi_new_2022}). The Cox et al. (\citeyear{cox_statistical_2020}) frequency-domain window-rejection algorithm (FDWRA) looked to accelerate the process of removing contaminated time windows leading to spurious peaks (recall Figure \ref{fig:1}c), a time consuming process if done manually. Cox et al. (\citeyear{cox_statistical_2020}) showed that their FDWRA performed well on a large dataset of 114 HVSR measurements from Wellington, New Zealand \citep{cox_dynamic_2018, vantassel_mapping_2018}. The FDWRA was subsequently implemented into \textit{hvsrpy} \citep{vantassel_jpvantasselhvsrpy_2020} and \textit{geopsy} \citep{wathelet_geopsy_2020} and used for other large HVSR datasets \citep{hallal_hv_2021}. However, these datasets included HVSR measurements with primarily one resonance and did not include a significant number of HVSR measurements with no or multiple resonances. As a result, the Cox et al. (\citeyear{cox_statistical_2020}) FDWRA still requires an analyst to go through manual resonance identification (recall Figure \ref{fig:1}d) before applying their algorithm. In addition, Cox et al. have repeatedly stated that the tuning parameter for their frequency-domain window-rejection algorithm (i.e., $n$, which represents the number of standard deviations from the median $f_{0}$ value), which is generally recommended as $n=2$, should be carefully considered and may need to be modified by the user on a case-by-case basis after scrutinizing the results \citep{cox_statistical_2020, vantassel_hvsrweb_2021}. The need to still perform resonance identification and the need for direct user-involvement to set an appropriate value for the method's hyperparameter limits some of the benefits of automating the window-rejection process. Therefore, while the Cox et al. (\citeyear{cox_statistical_2020}) FDWRA can help to accelerate the removal of outliers if the HVSR has only a single peak and the tuning parameter $n$ is selected carefully, it does not address the need for fully-automated resonance identification. The second algorithm, developed by Yazdi et al. (\citeyear{yazdi_new_2022}), looks at four different methods for identifying valid resonance peaks using a small dataset of 50 earthquake recordings assembled by Wang (\citeyear{wang_predictability_2020}). However, the methodologies proposed by Yazdi et al. (\citeyear{yazdi_new_2022}) do not address manual window rejection or manual resonance identification, the two most time-consuming aspects of manual HVSR processing. In summary, while the Cox et al. (\citeyear{cox_statistical_2020}) and Yazdi et al. (\citeyear{yazdi_new_2022}) semi-automated approaches seek to accelerate certain aspects of HVSR processing, they do not provide a fully-automated HVSR processing algorithm.

In contrast, this study presents AutoHVSR as a fully-automated algorithm capable of processing HVSR with zero, one, or multiple resonances and returning the uncertainty of each identified resonance and that of the mean/median HVSR curve. Its development involved three primary stages: dataset preparation, algorithm development, and algorithm evaluation. Dataset preparation involved aggregating a large dataset of 1109, 3-component microtremor recordings and processing them manually using best practices to establish a benchmark for later comparison. Algorithm development involved designing each component of the AutoHVSR algorithm to perform its associated subtask using a combination of traditional signal processing and state-of–the-art machine learning. Algorithm evaluation involved assessing the AutoHVSR algorithm in terms of its ability to replicate analyst-guided processing and accelerate the time to solution. Following development, the AutoHVSR algorithm was evaluated using the 1109-member dataset aggregated for its development and a separate 129-member dataset from Canterbury, New Zealand. Finally, while the AutoHVSR algorithm was originally developed using micromtremor measurements where the horizontal components were combined using the geometric mean, it is shown to generalize without modification to micromtremor measurements where the two horizontal components are rotated azimuthally and to eHVSR measurements. The AutoHVSR algorithm has been made publicly available in v0.3.0 of HVSRweb \citep{vantassel_hvsrweb_2021}, it is hosted on resources provided by DesignSafe-CI \citep{rathje_designsafe_2017}, and can be accessed at \href{https://hvsrweb.designsafe-ci.org/}{https://hvsrweb.designsafe-ci.org/}.

\section*{Dataset Preparation}

\subsection*{Microtremor Data}

The dataset used to develop the AutoHVSR algorithm includes a total of 1109, 3-component noise recordings collected over the prior six years. Note that this dataset is approximately 10 times larger than the dataset used by Cox et al. (\citeyear{cox_statistical_2020}) and approximately 22 times larger than the dataset used by Yazdi et al. (\citeyear{yazdi_new_2022}) for developing their semi-automated approaches. Measurements were made using either 3-component broadband seismometers, in particular 20s period or 120s period Nanometrics Trillium Compacts paired with either a Taurus or Centaur digitizer, or 3-component nodal stations, in particular a Magseis Fairfield 5 Hz ZLand 3C. Microtremor measurements were acquired under a variety of installation conditions, including complete burial, complete burial with weighted cover, partial burial, partial burial with weighted cover, and direct placement on top of hard ground with weighted cover. Furthermore, microtremor measurements were collected in a variety of geographic locations, including the continental United States (706), New Zealand (272), and East Asia (131), and in a variety of geologic settings, including soil deposits that range from very shallow (a few meters) to very deep (several thousands of meters) with subsurface stiffnesses that range from very soft (e.g., reclamation fill) to very hard (e.g., basalt outcrops). Recordings were made with a sampling rate of either 100 or 200 Hz for total durations ranging between 10 minutes and 17 hours. To avoid biasing the distribution of HVSR measurements toward those with longer recording times and to ensure the dataset was representative of recording times typically used for mHVSR measurements, all long ambient noise records were truncated to their first 180 minutes (3 hours). In summary, the assembled dataset exhibits diversity in terms of its acquisition conditions, geologic setting, and ambient noise field. Much of the data used for this study (\textgreater 40\%) has been made public as parts of prior dataset publications \citep{cox_dynamic_2018, vantassel_subsurface_2023}.

\subsection*{Traditional Processing}

All 1109, 3-component microtremor time-domain recordings were first prepared using a traditional processing workflow. To simplify the selection of appropriate processing parameters, in particular time window length, all microtremor recordings were divided into 35 time windows of equal length. The number 35 was selected such that after the window-rejection process, which typically removes between 5 and 10 windows, the remaining windows (25 to 30) would be sufficient for robust statistical calculations. The 3-components of each time window had their linear trend removed and a 5\% cosine taper (2.5\% off of both ends) applied. Each time window was transformed to the frequency domain, the horizontal components were combined using the geometric mean, and the vertical and horizontal components were smoothed using Konno and Ohmachi (\citeyear{konno_ground-motion_1998}) smoothing with a bandwidth ($b$) of 40. During smoothing, HVSR were resampled logarithmically between 0.05 and 50 Hz using 256 points. To ensure consistency between the time window length and minimum processing frequency, the frequency vector was truncated to those frequencies with at least 15 significant cycles. All HVSR calculations were performed using v1.0.0 of the open-source Python package \textit{hvsrpy} \citep{vantassel_jpvantasselhvsrpy_2021}.

\begin{figure}[!]
    \centering
	\includegraphics[width=1.0\textwidth]{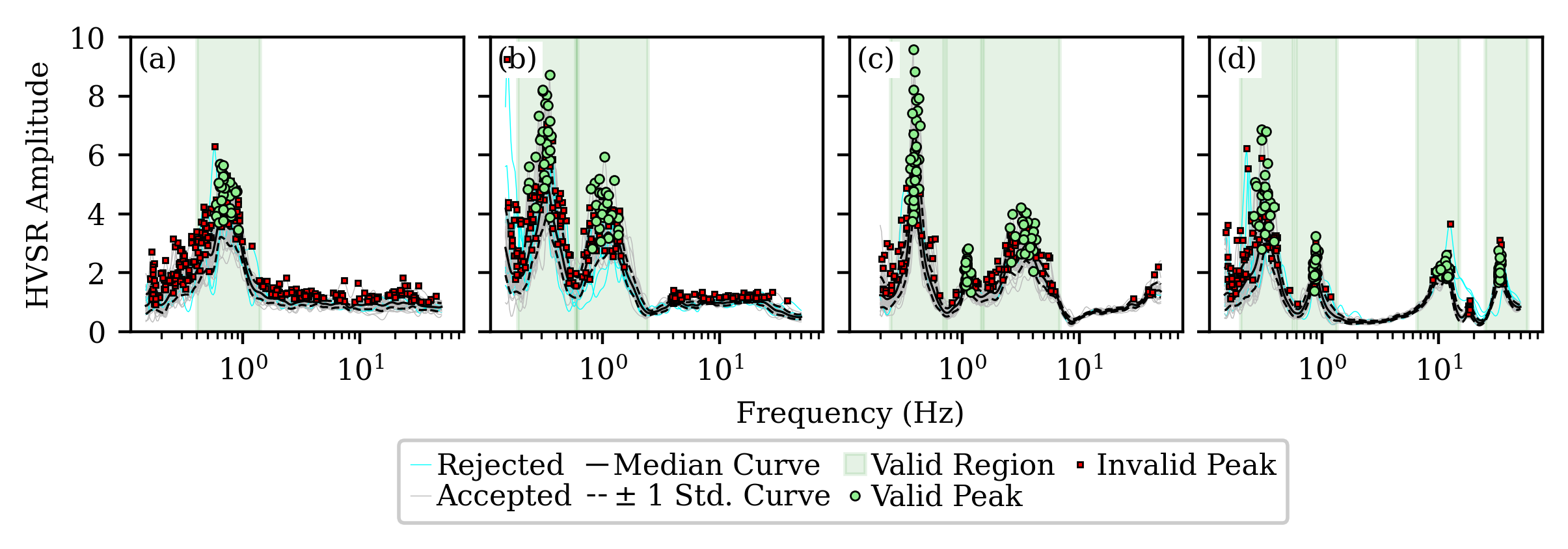}
	\caption{Summary of the dataset preparation procedure using four examples with (a) one, (b) two, (c) three, and (d) four clear HVSR resonances. Each panel includes the HVSR curves from the rejected time windows (thin light line), HVSR curves from the accepted time windows (thin dark line), lognormal median HVSR curve (dark solid line), and $\pm$ one lognormal standard deviation HVSR curve (dark dashed lines). In addition, each panel includes the regions manually identified through visual resonance identification (light fill), and the resulting valid and invalid peaks (light circles and dark squares, respectively).}
	\label{fig:2}
\end{figure}

Next, the calculated HVSRs were plotted for manual window rejection and resonance identification by a single analyst (i.e., the first author). Manual window rejection was performed first and involved removing any time window that was deemed to be an outlier in the HVSR domain (recall Figure \ref{fig:1}c). The window rejection process is subjective, but nonetheless critical for removing spurious HVSR curves resulting from time windows contaminated by non-ideal recording conditions (e.g., persons walking nearby). Following manual window rejection in the HVSR domain, each HVSR measurement was reprocessed for manual resonance identification (recall Figure \ref{fig:1}d). The HVSR measurements used in this study were identified to contain between zero and five significant resonances. Each identified HVSR resonance was manually defined by a lower and upper bounding frequency (recall Figure \ref{fig:1}e). Each identified HVSR resonance was evaluated using the SESAME clarity criteria \citep{sesame_guidelines_2004}; only those which passed the clarity criteria (5 or 6 out of 6) or those which nearly passed the clarity criteria (3 or 4 out of 6) but still visually appeared to be clear by the analyst were ultimately accepted. Following manual resonance identification, the data was reprocessed to identify all peaks in every HVSR time window using an automated peak finding routine. By a peak here we refer to any point on an HVSR curve whose neighboring points have a smaller value than itself. Peaks that were found inside of the upper and lower frequency bounds defined during manual resonance identification and that passed the FDWRA by Cox et al. (\citeyear{cox_statistical_2020}) with $n=2.5$ were marked as valid and all others (i.e., those not associated with any identified resonance) were marked as invalid. Figure \ref{fig:2}a, b, c, and d shows four different HVSR measurements with one, two, three, and four HVSR resonances, respectively. In each panel the HVSR curves from the accepted time windows are shown with a dark thin line, whereas the HVSR curves from the rejected time windows are shown in a light thin line. The lognormal median curve and its corresponding $\pm$ one lognormal standard deviation range computed from the accepted HVSR curves are shown with dark solid and dashed lines, respectively. The light shaded regions indicate the frequency range that was identified through manual resonance identification. The resulting valid and invalid peaks are shown with light circles and dark squares, respectively. The entire dataset of 1109 ambient noise recordings, includes 38,815 time windows and 285,441 potential peaks (approximately 8 peaks per window), of which 31,039 were labeled as valid (11\%) and 254,402 were labeled invalid (89\%) following the process described above. Importantly the data was processed ``blind'' using only the analyst's judgment and without any site-specific information, for example local geology, to constrain the manual window rejection and resonance identification processes. In this way, the data used to develop the AutoHVSR algorithm has been prepared in a manner suitable for future blind prediction. To process the entire 1109-member dataset using the traditional processing procedure took approximately 30 hours to complete.

\section*{Algorithm Development}

In broad terms, the AutoHVSR algorithm seeks to eliminate the need for direct analyst interaction when processing large HVSR datasets, while still producing robust HVSR results. In particular, the AutoHVSR algorithm seeks to automate two main time-consuming tasks: (1) the removal of low-quality time windows (i.e., manual window rejection, recall Figure \ref{fig:1}c) and (2) the identification of HVSR resonances (i.e., manual resonance identification, recall Figure \ref{fig:1}d). Therefore, the AutoHVSR algorithm, shown schematically in Figure \ref{fig:3}, if treated purely as an opaque box, accepts time-domain microtremor (or as demonstrated later, earthquake) recordings and user-defined processing parameters as input (Figure \ref{fig:3}a) and returns meaningful statistical measures (mean/median and uncertainty) for the full HVSR curve and each significant resonance (Figure \ref{fig:3}g). To accomplish this, the AutoHVSR algorithm combines traditional signal processing tools with state-of-the-art machine learning models to attain human-level performance on the task of HVSR processing. It is important to note that the AutoHVSR algorithm does not completely remove the need for an experienced analyst. Instead, it concentrates the analyst's time on the novel aspects of HVSR analysis, which includes interpreting HVSR resonances with regard to the site's geologic settings, by minimizing the analyst's time on tedious data preparation and processing tasks. The details of the main steps within the AutoHVSR algorithm are summarized in the following sections. 

\subsection*{HVSR Calculation}

\begin{figure}[t!]
    \centering
	\includegraphics[width=1.0\textwidth]{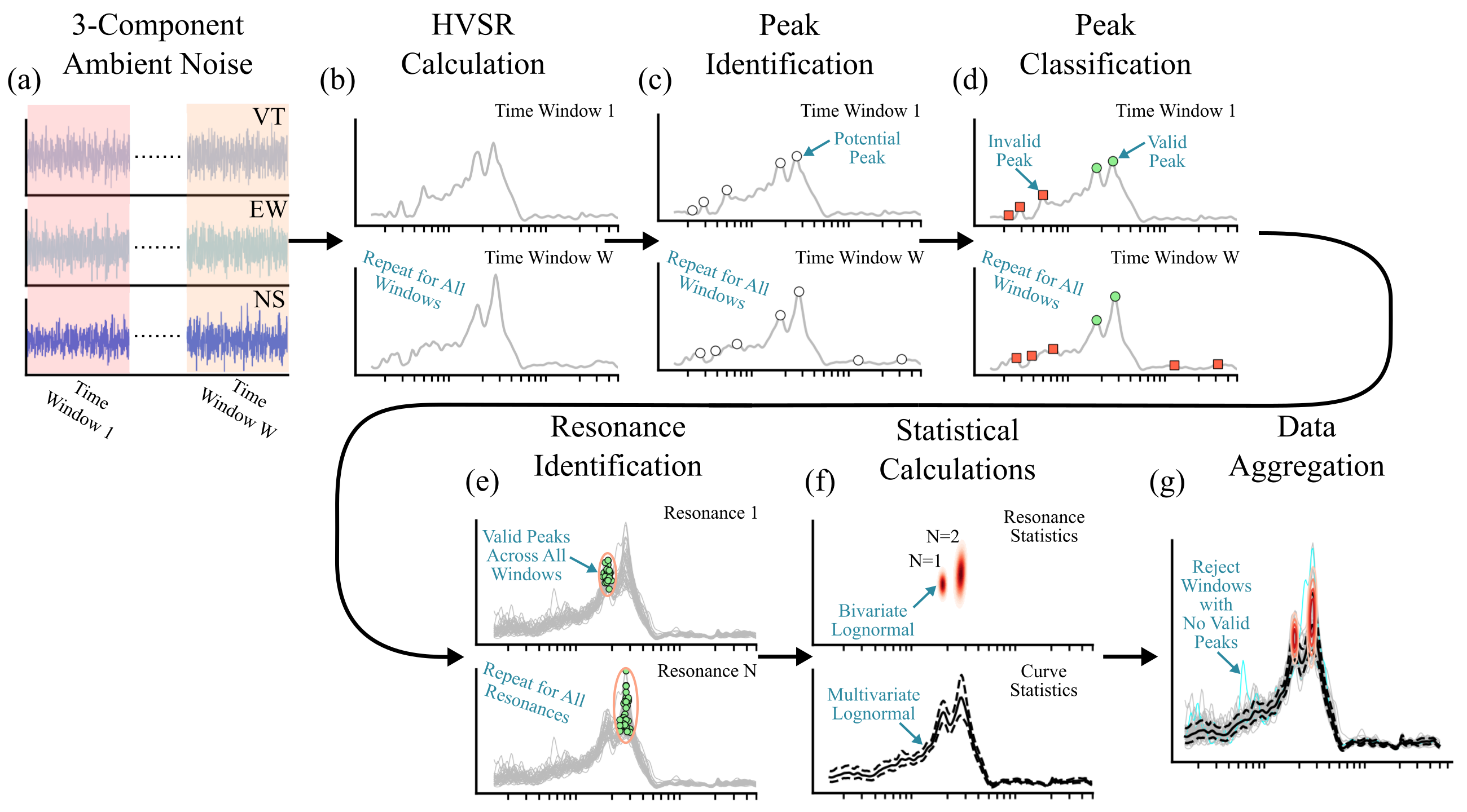}
	\caption{Summary of the AutoHVSR workflow from (a) the 3-component microtremor time-domain recording split into $W$ windows, (b) HVSR calculations, (c) peak identification, and (d) peak classification for each window, (e) resonance identification via the clustering of valid peaks into $N$ resonances, (f) statistical calculations on each identified resonance and the full HVSR curve, and (g) data aggregation. Note that while the method is illustrated using windowed microtremor measurements (i.e., mHVSR) the method equally applies to HVSR computed using a set of earthquake recordings (i.e., eHVSR).}
	\label{fig:3}
\end{figure}

The HVSR calculation component of the AutoHVSR algorithm takes time-domain recordings and user-defined processing parameters as input and returns the associated HVSR curves, see Figure \ref{fig:3}a. As the HVSR processing approach has been discussed previously, it will not be repeated here.

\subsection*{Peak Identification}

Ideally, the peak identification component of the AutoHVSR algorithm would accept an HVSR curve from an individual time window and identify the location of all significant peaks. However, this task is quite challenging for traditional algorithms due to the subjectivity implied by the word ``significant''. Stated differently, while finding all the peaks in a given curve is trivial (i.e., finding all the points where the points to its left and right have a smaller value), finding only those which are ``significant'' is not trivial due to the difficulty of translating the significance requirement into logical statements comprehensible by a computer (i.e., into computer code). Therefore, the task of finding all of the significant peaks was divided into two algorithm components: peak identification via a rigorous peak-finding routine and peak classification via a machine learning classifier capable of capturing the subjectivity implied by the term ``significant''.

The peak identification routine, shown schematically in Figure \ref{fig:3}c, was selected with two main criteria in mind. First, it should not fail to identify any peak deemed significant. That is, the routine should be capable of identifying a set of possibly-significant peaks that is a superset of the set of significant peaks. Second, it should include appropriate minimum thresholds such that it does not select all available peaks, as selecting all available peaks could pose unnecessary computational demands on later components of the algorithm. Importantly, since many previous peak-finding algorithms have been developed and made publicly available, this study chose to adopt those implemented in the aptly named $find\_peaks$ function of the open-source Python package \textit{scipy} \citep{virtanen_scipy_2020}. The $find\_peaks$ function, in addition to the data to be searched, accepts eight input arguments. These eight function arguments control the user's ability to define the properties of a peak and, as a result, filter the returned results. For the present study, three of these parameters were deemed of interest: $threshold$, $distance$, and $prominence$. The $threshold$ parameter returns only those peaks whose amplitude are greater than the minimum specified and less than the maximum specified. Minimum $threshold$ values of 0.0 and 2.0 were considered during hyperparameter selection. Note that no maximum $threshold$ was considered. The $distance$ parameter returns only those peaks whose minimal horizontal distance is greater than the integer specified, effectively requiring that frequency peaks occur some minimum distance from one another. $distance$ values of 1, 3, 5, and 7, which correspond approximately to distances of 0.012, 0.035, 0.059, and 0.082 in log(Hz), were considered during hyperparameter selection. The $prominence$ parameter effectively measures the peak's height with respect to the closest baseline (i.e., its closest local minimum), and therefore, unlike the two prior parameters, is a relative measure. $prominence$ values of 0.0, 0.25, 0.5, 1.0, and 2.0 were considered during hyperparameter selection. Following hyperparameter selection, the use of a minimum $prominence$ of 0.25 and declining to set minimums for $threshold$ and $distance$ was found to provide a good trade-off between the two aforementioned criteria of identifying all valid peaks while not observing too many invalid peaks. The ratio of invalid to valid peaks across the entire dataset was approximately 8.2 to 1.

\subsection*{Peak Classification}

The peak classification component accepts the potential peaks identified previously and classifies each peak as significant or not significant (see Figure \ref{fig:3}d). As discussed previously, this task requires the subjective evaluation of the term significant, making the utilization of a purely rule-based algorithm challenging to develop. A useful alternative for solving perceptual problems (i.e., problems that are easy for humans to reason about but difficult to implement in strict mathematical statements) is machine learning \citep{goodfellow_deep_2016}. The area of machine learning includes many algorithms, however, one which has performed particularly well, as evidenced by its good performance in machine learning competitions \citep{chollet_deep_2021}, is gradient boosting \citep{friedman_greedy_2001, friedman_stochastic_2002}, in particular, gradient-boosted decision trees. The gradient-boosted decision trees algorithm is similar in some regards to the popular random forest algorithm \citep{tin_kam_ho_random_1995}, where many decision trees that have each been developed on random subsets of the data are combined to create a single prediction algorithm. The main difference between the random forest and gradient-boosted trees algorithms are that while each tree in the random forest algorithm is built from scratch, the trees in the gradient-boosted algorithm focus on improving the predictions made by the previous collection of trees.

Applying the gradient-boosted trees algorithm to the problem of peak classification requires two main tasks: identification and extraction of relevant features for prediction and hyperparameter tuning. Feature identification, also known as feature engineering, is the process of identifying properties of the data from the training set that may be indicative of the response variable. In this case, 23 features (or predictive variables) were identified for representing each potential peak:

\begin{itemize}
    \item peak frequency,
    \item peak amplitude,
    \item peak prominence,
    \item amplitude of the mean curve at the peak frequency,
    \item standard deviation of the mean curve at the peak frequency,
    \item eight features to describe the HVSR's curves general shape,
    \item five features to describe the peak density in terms of frequency, and
    \item five features to describe peak density in terms of amplitude.
\end{itemize}

The response variable was a Boolean that denoted whether a peak was valid (True) or invalid (False). For developing the gradient-boosted tree algorithm, this study used the Python package of the open-source Extreme Gradient Boosting (xgboost) software \citep{chen_xgboost_2016}. For hyperparameter tuning, two hyperparameters of the gradient-boosted model were held fixed and the others systematically varied to assess their effect on algorithm performance. The two parameters held constant were the gradient scale on the positive classes, set to 8.2, and the fraction of the total data provided for each tree's development, set at 1. The five varied parameters were the learning rate, number of trees, maximum tree depth, minimum loss reduction for split, and L2 regularization on the model's weights. The set of values considered for each hyperparameter are listed in Table \ref{table:1}. Hyperparameter values were tuned by using a grid search across the 9072 potential hyperparameter combinations, where each combination was rigorously assessed using stratified 5-fold cross validation. The best performing model was defined by a learning rate of 0.1, an ensemble of 500 trees, maximum depth of 10, minimum loss reduction for split of 0.1, and L2 regularization on the model's weights of 10.0. Following hyperparameter selection, the final gradient-boosted tree model was trained using the entire training set and attained a log loss of 0.017, F1 score (the harmonic mean of precision and recall) of 0.97, and an area under the receiver operating characteristics curve (ROC AUC) of 0.99, indicating acceptable performance on the training data. The fully trained model was then evaluated on the testing set, where it attained a log loss of 0.077, F1 score of 0.88, and a ROC AUC of 0.99, indicating good performance as a predictive model.

\begin{table}[b]
\centering
\caption{Considered and selected hyperparameter values for the gradient-boosted trees algorithm used as the peak classification component of AutoHVSR. Hyperparameter values were selected via grid search across all 9072 potential combinations.}
\label{table:1}
\begin{tabular}{@{}ccc@{}}
\toprule
\textbf{Hyperparameter}          & \textbf{Considered Values}          & \textbf{Selected Value} \\ \midrule
Learning Rate                    & 0.03, 0.1, 0.3, 0.5                 & 0.1                     \\
Number of Trees                  & 30, 50, 100, 300, 500, 1000         & 500                     \\
Maximum Tree Depth               & 1, 2, 4, 8, 10, 12                  & 10                      \\
Minimum Loss Reduction for Split & 0.0, 0.03, 0.1, 0.3, 1.0, 3.0, 10.0 & 0.1                     \\
L2 Regularization on the Model's Weights & 0.0, 0.03, 0.1, 0.3, 1.0, 3.0, 10.0, 30.0, 100.0 & 10 \\ \bottomrule
\end{tabular}
\end{table}

\subsection*{Resonance Identification}

The resonance identification component accepts the peaks that have previously been classified as valid and groups them according to their proximity to identify HVSR resonances (Figure \ref{fig:3}e). Multiple alternatives were considered for the resonance clustering component, including clustering using k-means \citep{macqueen_methods_1967}, fitting of a Gaussian mixture model (GMM), and density-based spatial clustering of applications with noise (DBSCAN) \citep{ester_density-based_1996}. While k-means performed well on some limited trial examples, it did not perform well when the number of clusters ($k$) could not be selected manually. Fitting a GMM, while not strictly clustering, would in effect allow the final two steps of the algorithm (i.e., resonance clustering and statistical calculations) to be combined into a single step. This approach was particularly appealing given that HVSR peak frequency and amplitude is commonly modeled as lognormally distributed. Therefore, by fitting a GMM to the log-transformed frequencies and amplitudes this domain-specific knowledge could be incorporated into the resonance clustering process. However, when the GMM was applied to the training dataset it performed poorly. This was unexpected, however, a statistical analysis of the HVSR resonances identified through traditional processing found strong evidence (p-value \textless 0.01), using the test developed by Shapiro and Wilk (\citeyear{shapiro_analysis_1965}), that 371 of the 1181 (31\%) resonances identified were not lognormally distributed in terms of frequency. Ultimately, these shortcoming of k-means and GMM led to the utilization of the DBSCAN algorithm, as it does not require the number of clusters to be known a priori (an improvement over k-means), does not require the data to be lognormally distributed (an improvement over GMM), and performs well in the presence of noise even when the sample size is small (\textless 30).

The application of the DBSCAN algorithm required first identifying those features from each peak that should be clustered and second selecting appropriate algorithm hyperparameters. Regarding the data that should be clustered, four features were of interest: $peak\ frequency$, $peak\ amplitude$, $peak\ prominence$, and $amplitude\ of\ the\ mean\ curve\ at\ the\ peak\ frequency$. In addition to the combination of these parameters, each could also be log-transformed to potentially further optimize the feature set. Given the number of possible combinations, an exhaustive search was not deemed reasonable given the computational expense of evaluating the quality of each combination (discussed in detail later). Therefore, only the following four combinations were considered where only $peak\ frequency$ was log-transformed: $peak\ frequency$, $peak\ frequency$ and $peak\ amplitude$, $peak\ frequency$ and $peak\ prominence$, and $peak\ frequency$ and $amplitude\ of\ the\ mean\ curve\ at\ the\ peak\ frequency$. For each data combination, the hyperparameters were tuned using a grid search. The hyperparameters that were tuned included the maximum distance between two points for them to be considered in one another's neighborhood ($eps$) and the minimum number of points required to form a set of core points ($min\_samples$). In addition, for the latter three feature combinations that include two input features, an additional hyperparameter was considered to scale the second parameter relative to the first (i.e., $peak\ frequency$). The hyperparameter values considered are listed in Table \ref{table:2}.

\begin{table}[]
\centering
\caption{Considered and selected hyperparameter values for the density-based spatial clustering of applications with noise (DBSCAN) algorithm used as the resonance identification component of AutoHVSR. Hyperparameter values were selected via grid search across all 72 potential combinations.}
\label{table:2}
\begin{tabular}{@{}ccc@{}}
\toprule
\textbf{Hyperparameter}                  & \textbf{Considered Values}     & \textbf{Selected Value} \\ \midrule
Neighborhood Distance                    & 0.1, 0.2, 0.3                  & 0.2                     \\
Minimum Points Needed to Form a Core Set & 5, 10, 15, 20                  & 10                      \\
Secondary Feature Scaling Parameter      & 0.0, 0.25, 0.50, 1.0, 2.0, 4.0 & 0.0                     \\ \bottomrule
\end{tabular}
\end{table}

While clustering is generally considered an unsupervised learning method, here we use it as if it were a supervised learning method for the purposes of selecting hyperparameters. To use DBSCAN as if it were a supervised learning method involved comparing the class predictions provided by DBSCAN for each peak against those in the training set, which here acts like a large validation set, as no parameters are being “learned” directly from the data. The main challenge of using DBSCAN in a supervised way is that the classifications provided for each peak can be any integer between -1 and N-1, where -1 designates noise and a positive integer 0 to N-1 designates one of N non-noise clusters. In contrast, the dataset previously developed classifies each peak to integers between -1 and 4 for noise and five possible resonant frequencies, respectively. To determine an appropriate metric to evaluate the clustering quality, after each application of DBSCAN an iterative comparison was performed to compare the N potential class designations from DBSCAN with the 6 potential class designations identified in the training data. Because this search requires the consideration of every permutation of the two class designation arrays, the process is computationally demanding, especially when DBSCAN produces a large number of clusters. To avoid undue computational cost three strategies were used: reducing the number of feature combinations considered (previously discussed), avoiding small values of $eps$, and avoiding small values of $min\_samples$. The best performing combination was utilizing frequency data as input, with $eps=0.2$, and $min\_samples=10$, which attained an F1 score of 0.84 on the training set. Note the log loss and ROC AUC are not provided here as they cannot be computed using the outputs provided by DBSCAN.

While an F1 score of 0.84 is indicative of good performance, an investigation was performed to better understand what types of resonances were being misclassified. Through this investigation, the majority of misclassified resonances were found to occur for HVSR curves where two distinct peaks occurred in close proximity to one another and which were being incorrectly clustered together by the DBSCAN algorithm. Several methods were explored to remedy this issue, including decreasing the DBSCAN $eps$ variable, using k-means clustering with $k=2$ on each high variance cluster identified by DBSCAN, and splitting high variance clusters at their median or mean value. Ultimately, splitting large clusters at their mean value was found to perform the best of the methods considered. The addition of the splitting process improved the performance from an F1 score of 0.84 to 0.88 on the training set, a significant improvement. When applied to the testing set the resonance identification algorithm achieved an F1 score of 0.88. Throughout the development of the resonance identification algorithm the peaks used for clustering were those identified using the gradient-boosted peak algorithm previously discussed. If instead the resonance identification algorithm is performed directly on the true peak classifications, and thereby more directly assesses the error due to the resonance identification algorithm alone, the F1 score on the training and testing sets increase to 0.98 and 0.96, respectively, indicating good predictive performance by the resonance identification algorithm.

\subsection*{Statistical Calculations}

The final component of the AutoHVSR algorithm is the statistical calculations (see Figure \ref{fig:3}f). This component accepts two primary inputs, the clustered resonances and the HVSR curves from each time window, for computing resonance and curve statistics, respectively. For each resonance we calculate the mean and standard deviation for frequency and amplitude assuming a bivariate lognormal distribution. The statistical calculations are repeated for all N resonances identified. To calculate the statistics for the HVSR curve a few options are available. First, and most commonly, all HVSR curves are incorporated into the statistical calculation. This option, while commonly done, can include time windows whose HVSR is inconsistent with the majority of HVSR measurements and are thereby circumspect. Others have proposed to remove those windows whose peaks are inconsistent with the fundamental resonant frequency \citep{sesame_guidelines_2004, cox_statistical_2020}. However, doing so is not possible if the HVSR curve does not have a clear fundamental resonant frequency and less meaningful if it contains multiple clear resonances. In the present study, we opt to permit any HVSR curve that has at least one clear peak and reject any HVSR curve that has no clear peak, unless no clear resonances have been identified across all HVSR time windows in which case all curves are maintained. The mean and standard deviation of the accepted HVSR curves are then computed assuming a multivariate lognormal distribution.

\section*{Algorithm Evaluation}

\begin{figure}[t!]
    \centering
	\includegraphics[width=0.5\textwidth]{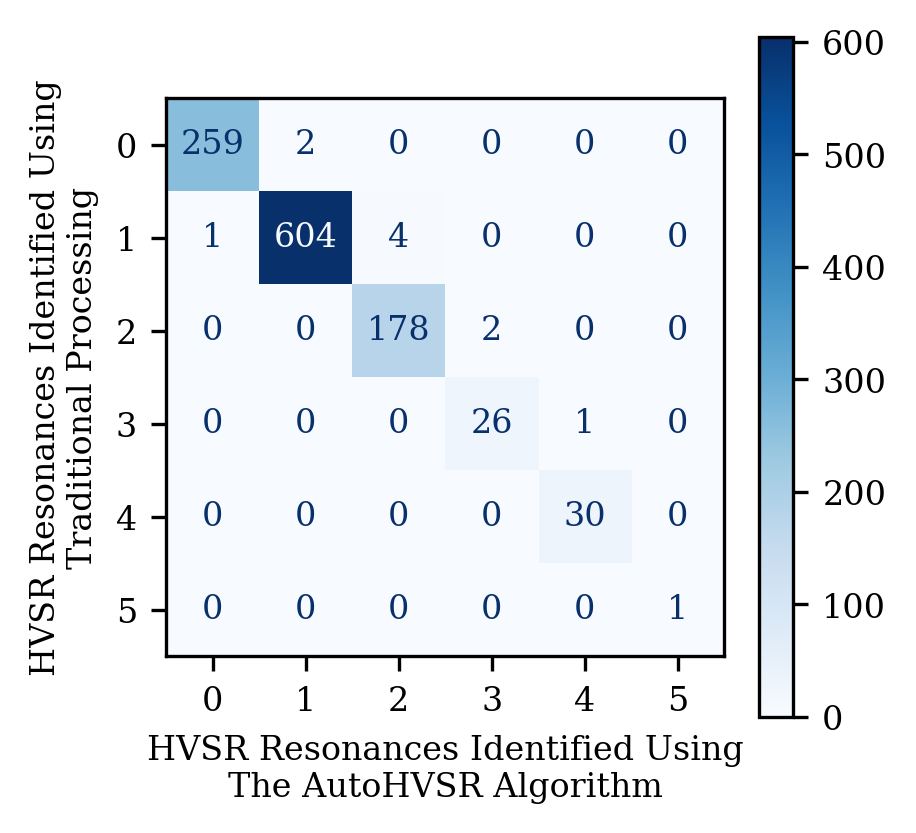}
	\caption{Confusion matrix comparing the number of resonances identified in an HVSR measurement from the AutoHVSR algorithm and traditional processing. The main diagonal indicates the quantity of HVSR measurements that had their number of resonances correctly identified, these include the vast majority of measurements in the assembled dataset (i.e., 1099 of the 1109 or 99.1\%).}
	\label{fig:4}
\end{figure}

\begin{figure}[t!]
    \centering
	\includegraphics[width=0.75\textwidth]{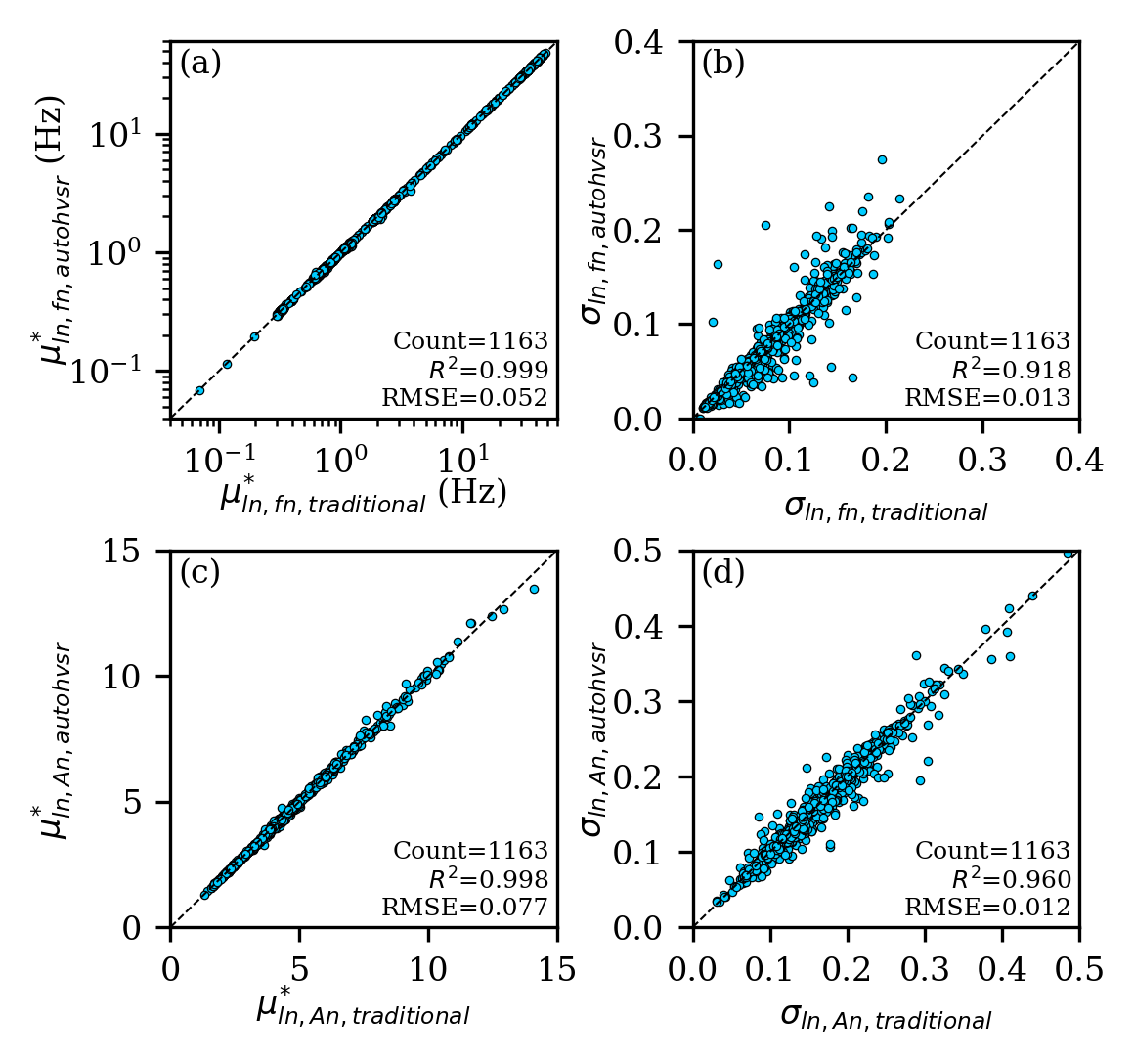}
	\caption{Assessment of the AutoHVSR algorithm's performance in terms of the (a) exponentiated lognormal mean frequency ($\mu_{ln,fn}^*$), (b) lognormal standard deviation of frequency ($\sigma_{ln,fn}$), (c) exponentiated lognormal mean amplitude ($\mu_{ln,An}^*$), and (d) lognormal standard deviation of amplitude ($\sigma_{ln,An}$) for the resonances identified by AutoHVSR and traditional processing.}
	\label{fig:5}
\end{figure}

In the previous sections the components of the AutoHVSR algorithm have been evaluated using the training and testing data. In this section the integration of these components is evaluated on the entire dataset. More specifically, the AutoHVSR algorithm is evaluated in terms of its ability to accurately determine the number of resonances in an HVSR measurement, the reliability of the AutoHVSR algorithm to capture the resonance's appropriate frequency and amplitude statistics, and the time to solution. Figure \ref{fig:4} presents a confusion matrix of the resonances identified in each HVSR measurement by the AutoHVSR algorithm and traditional processing. The values along the diagonal indicate the measurements that have had the correct number of resonances identified by the AutoHVSR algorithm, that is, the number of resonances determined by the AutoHVSR algorithm match the number of resonances determined through traditional processing. Figure \ref{fig:4} shows that the vast majority (i.e., 1099 of the 1109 or 99.1\%) of measurements are correctly classified. Furthermore, the majority (9 out of 10) of those that are misclassified exist in the upper triangle of the confusion matrix, indicating that the number of resonances identified by the AutoHVSR algorithm exceeds those identified by traditional processing. A review of these misclassified cases show that the majority (8 out of 9) of these have correctly identified the same peaks as the traditional processing procedure but also include an additional peak that could be classified as clear, but was not during traditional processing based on the SESAME criteria or the analyst's judgment. Next, to evaluate the quality of the resonances recovered by the AutoHVSR algorithm, Figure \ref{fig:5} compares the exponentiated lognormal mean frequency ($\mu_{ln,fn}^*$), lognormal standard deviation of frequency ($\sigma_{ln,fn}$), exponentiated lognormal mean amplitude ($\mu_{ln,An}^*$), and lognormal standard deviation of amplitude ($\sigma_{ln,An}$) from AutoHVSR and traditional processing for all HVSR measurements, where both methods predicted the same number of resonances. Excellent agreement is observed in terms of frequency and amplitude, as evidenced by high coefficient of determination ($R^2$) (above 0.998) and low root mean squared error (RMSE) (below 0.077). A comparison of the lognormal standard deviations also demonstrates high $R^2$ values (above 0.91) and low RMSE (below 0.013). Finally, in terms of time to solution the AutoHVSR algorithm took 13 minutes to transform the datasets 1109 ambient noise recordings into the HVSR results presented. When compared to the 30 hours require to process the entire 1109 member dataset using the traditional approach, the AutoHVSR algorithm provides a speed up of approximately 138.

\begin{figure}[t!]
    \centering
	\includegraphics[width=1.0\textwidth]{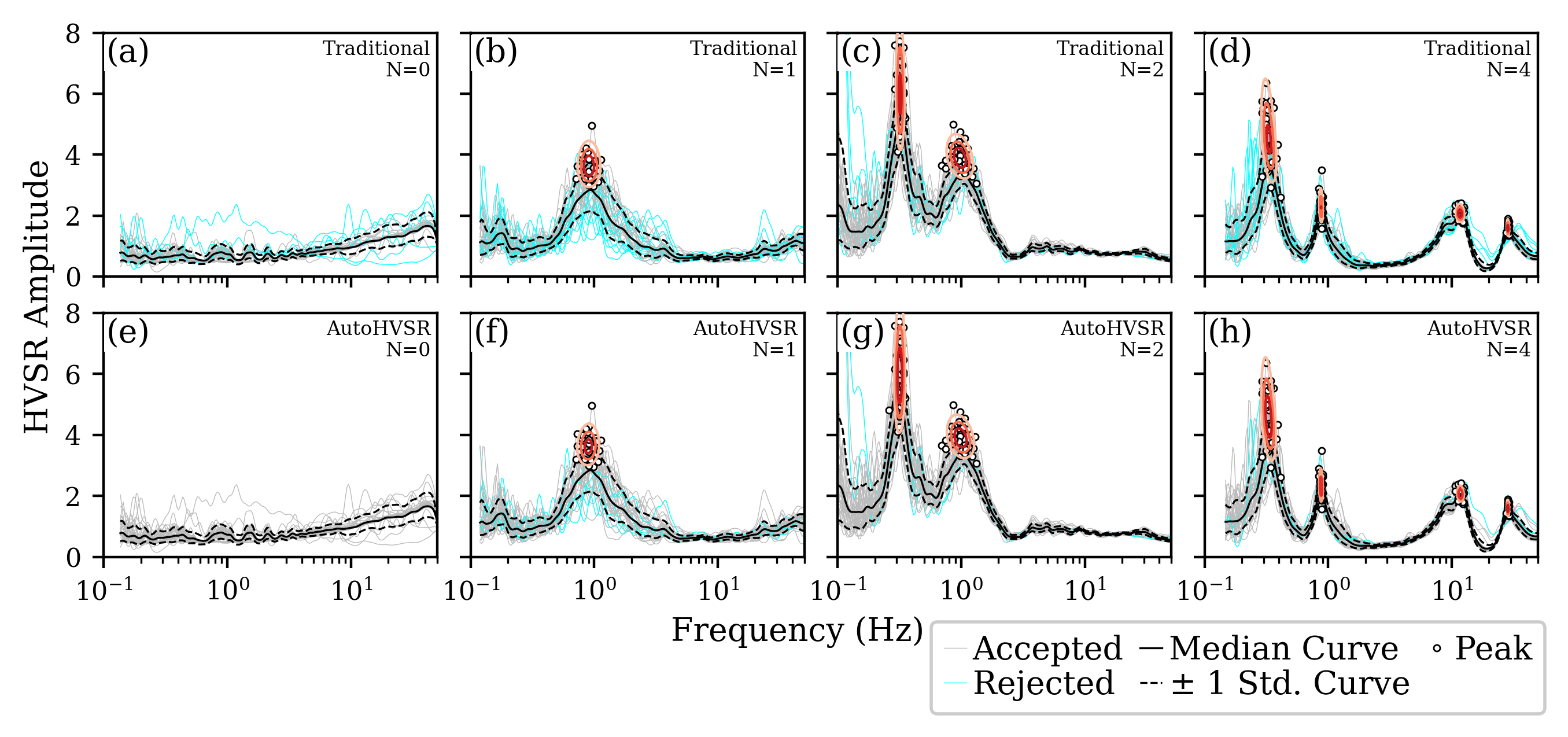}
	\caption{Examples from the testing set prepared using traditional processing (top row) and the AutoHVSR algorithm (bottom row). The four examples are taken from the main diagonal of the confusion matrix (Figure \ref{fig:4}) where the number of resonances predicted by AutoHVSR and traditional processing agree. These examples were selected as they contain (a \& e) zero, (b \& f) one, (c \& g) two, and (d \& h) four clear resonances (i.e., $N=$ 0, 1, 2, and 4, respectively). Each panel includes HVSR curves from the rejected time windows (thin light line), HVSR curves from the accepted time windows (thin dark line), lognormal median HVSR curve (dark solid line), and $\pm$ one lognormal standard deviation HVSR curve (dark dashed lines). The distribution of each significant resonance's frequency and amplitude are also shown.}
	\label{fig:6}
\end{figure}

\begin{figure}[t!]
    \centering
	\includegraphics[width=1.0\textwidth]{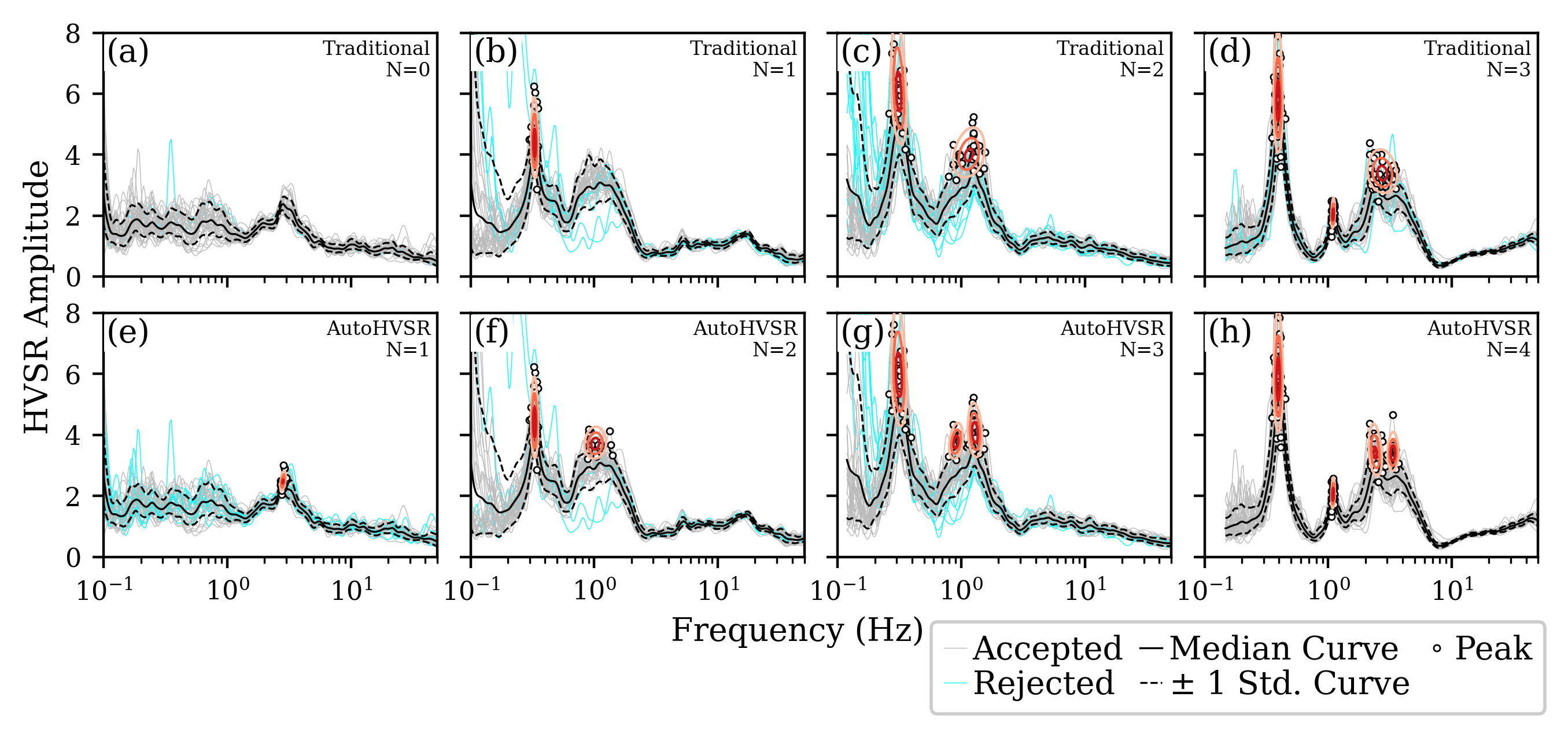}
	\caption{Examples from the testing set prepared using traditional processing (top row) and the AutoHVSR algorithm (bottom row). The four examples are taken from off the main diagonal of the confusion matrix (Figure \ref{fig:4}) where the number of resonances predicted by AutoHVSR and traditional processing disagree. These examples were selected as they illustrate that in general the AutoHVSR algorithm identified more resonances than those using traditional processing. Each panel includes HVSR curves from the rejected time windows (thin light line), HVSR curves from the accepted time windows (thin dark line), lognormal median HVSR curve (dark solid line), and $\pm$ one lognormal standard deviation HVSR curve (dark dashed lines). The distribution of each significant resonance's frequency and amplitude are also shown.}
	\label{fig:7}
\end{figure}

To demonstrate the effectiveness of the AutoHVSR algorithm more visually, Figures \ref{fig:6} and \ref{fig:7} present HVSR measurements drawn from the testing set. Figure \ref{fig:6} illustrates four examples representative of HVSR measurements from the main diagonal of the confusion matrix (Figure \ref{fig:4}) where the number of resonances predicted by AutoHVSR match those from traditional processing. The top row of Figure \ref{fig:6} illustrates HVSR with: (a) zero, (b) one, (c) two, and (d) four clear resonances identified through traditional processing. The bottom row [i.e., panels (e) through (f)] illustrates the same four examples as in the row above, but instead processed using AutoHVSR. Shown in each panel are the HVSR curves from the rejected time windows (thin light line), HVSR curves from the accepted time windows (thin dark line), lognormal median HVSR curve (solid dark line), and $\pm$ one lognormal standard deviation HVSR curves (dashed dark lines). The distribution of each resonance's frequency and amplitude are also shown. Good agreement is observed when comparing the window rejection and resonance identification using the traditional approach (top row of Figure \ref{fig:6}) and that achieved by AutoHVSR (bottom row of Figure \ref{fig:6}). As a consequence, the statistics on the mean curve and each identified resonance are also in good agreement. Figure \ref{fig:7} illustrates examples representative of HVSR measurements from off the main diagonal of the confusion matrix (Figure \ref{fig:4}) where the number of resonances predicted by AutoHVSR do not match those from traditional processing. Similar to Figure \ref{fig:6}, the top row of Figure \ref{fig:7} illustrates traditional processing and the bottom row processing with AutoHVSR. The examples shown are representative of the case where the number of resonances identified by the AutoHVSR algorithm exceeds those identified by traditional processing. These constitute the majority (8 out of 9) of the incorrect predictions and clearly show that AutoHVSR has correctly identified the same peaks as the traditional processing but has also identified additional peaks that could be classified as clear, but were not during traditional processing based on the SESAME criteria and/or the analyst's judgment.

\section*{Application to Canterbury, New Zealand Dataset}

The AutoHVSR algorithm was applied to a previously analyzed dataset from Canterbury, New Zealand. The dataset was prepared by various analysts over a period of six years. Note that the first author who prepared the 1109-member dataset aforementioned for developing AutoHVSR did not assist in the preparation of the Canterbury dataset, and therefore the analysts for both datasets are disjoint. Furthermore, in anticipation of using this dataset for assessing the robustness of the AutoHVSR algorithm, no HVSR measurements from Canterbury, New Zealand were included in the original 1109-member AutoHVSR dataset to avoid any potential region-specific information leakage. The processing of the Canterbury, New Zealand  dataset was performed using the H/V toolbox in the open-source software \textit{geopsy} \citep{wathelet_geopsy_2020} using processing parameters at the discretion of the analyst and which, therefore, varied on a measurement-by-measurement basis. The HVSR resonances in the Canterbury, New Zealand dataset were identified manually with the assistance of local geology and with information of nearby resonances identified from other nearby HVSR testing locations. Therefore, we consider the traditional processing of the Canterbury, New Zealand dataset ``geologically informed'', whereas the AutoHVSR algorithm analysis is performed completely ``blind''. These considerations make the Canterbury, New Zealand dataset a particularly challenging test case for the AutoHVSR algorithm. It is estimated that the traditional processing of the Canterbury, New Zealand dataset took approximately 4 hours.

\begin{figure}[t!]
    \centering
	\includegraphics[width=0.5\textwidth]{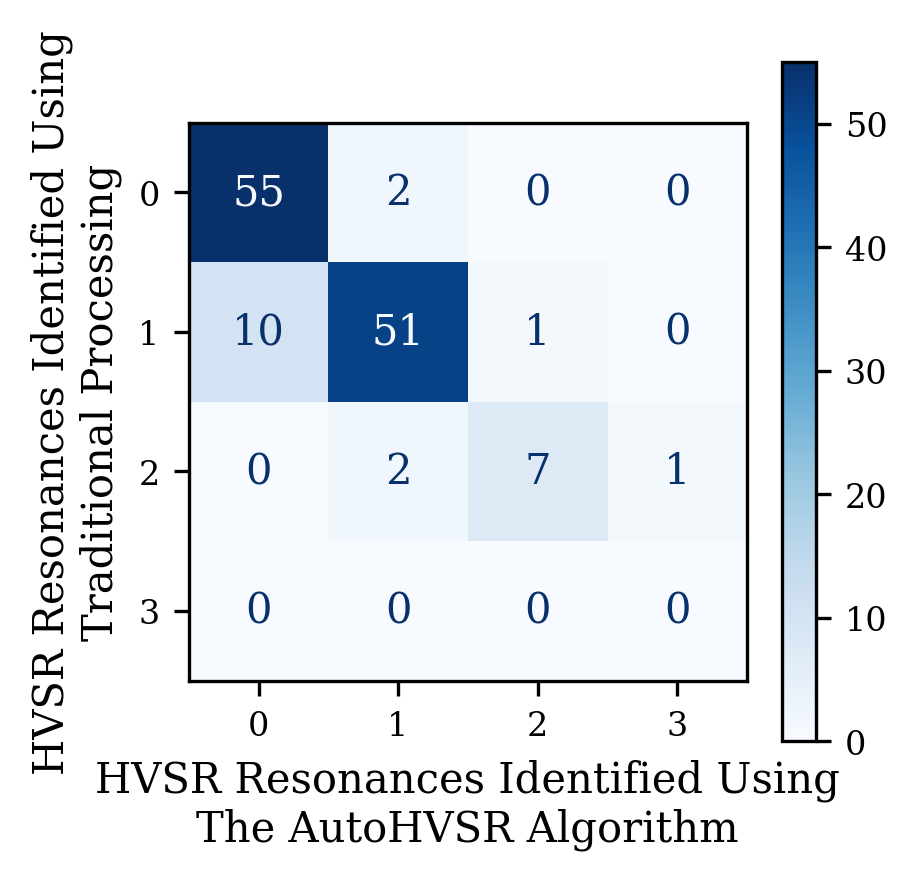}
	\caption{Confusion matrix comparing the number of resonances identified in an HVSR measurement from the AutoHVSR algorithm and traditional processing on the Canterbury, New Zealand dataset. The main diagonal indicates the quantity of HVSR measurements that had their number of resonances correctly identified, these include the vast majority of measurements in the assembled dataset (i.e., 113 of 129 or 88\%).}
	\label{fig:8}
\end{figure}

\begin{figure}[t!]
    \centering
	\includegraphics[width=0.75\textwidth]{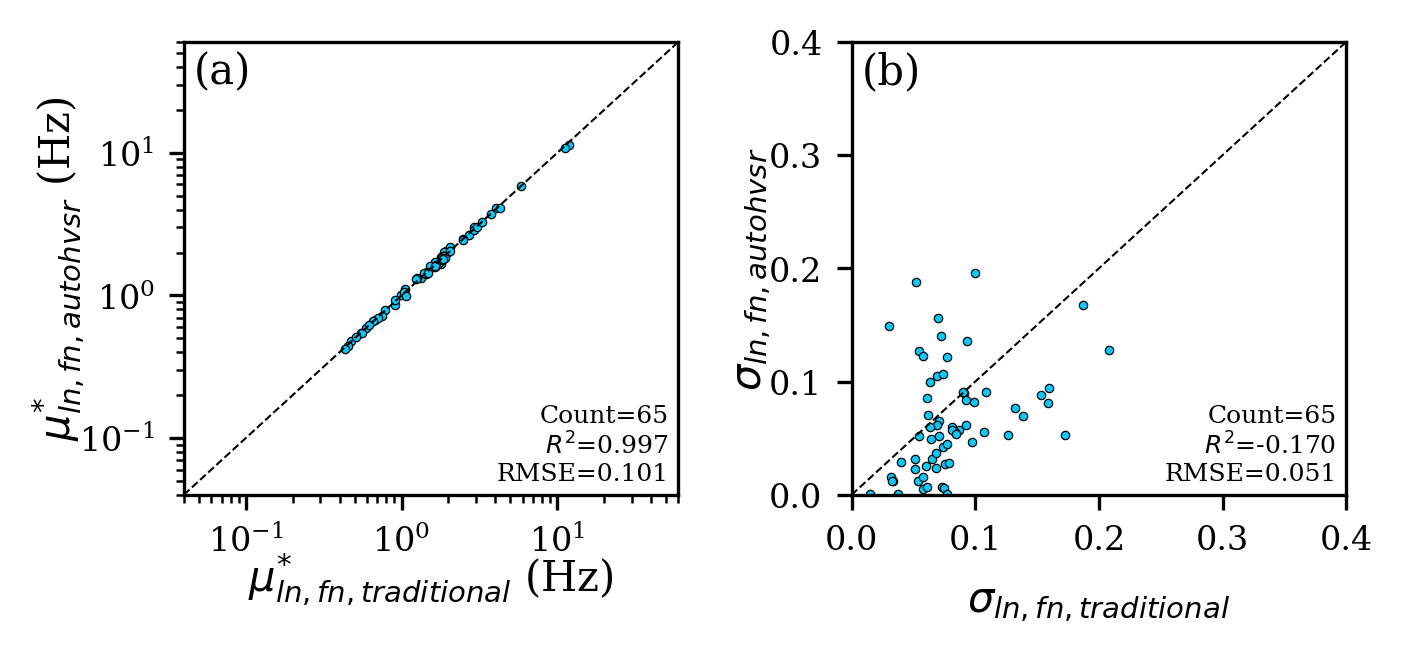}
	\caption{Assessment of the AutoHVSR algorithm's performance on the Canterbury, New Zealand dataset in terms of the (a) exponentiated lognormal mean frequency ($\mu_{ln,fn}^*$) and (b) lognormal standard deviation of frequency ($\sigma_{ln,fn}$) for the resonances identified by AutoHVSR and traditional processing. Note that amplitude information was not collected as part of the original analysis of the Canterbury, New Zealand dataset and therefore no comparisons are able to be made here.}
	\label{fig:9}
\end{figure}

The AutoHVSR algorithms performance is assessed via a confusion matrix in Figure \ref{fig:8} and regression plots in Figure \ref{fig:9}. The confusion matrix in Figure \ref{fig:8} assesses the number of resonances identified by the geologically-informed traditional processing and the blind AutoHVSR processing. The main diagonal indicates the number of HVSR measurements that had their number of resonances correctly identified, these include the vast majority of measurements in the assembled dataset (i.e., \textgreater87\% or 113 of 129). As with the 1109-member dataset, a majority of the off-diagonal cases (i.e., where the number of resonances identified by traditional and AutoHVSR processing disagree) are located in the upper right triangle of the confusion matrix, indicating that the AutoHVSR algorithm tended to identify more resonances than were identified manually. In general, these cases identified the same resonances as traditional processing but also identified additional resonances that were not identified during traditional processing. For those resonances that were identified by both AutoHVSR and traditional processing, Figure \ref{fig:9}a and \ref{fig:9}b compares the exponentiated lognormal mean frequency and lognormal standard deviation of frequency, respectively. Note that because the traditional processing was performed in \textit{geopsy}, which does not allow for the calculation of lognormal statistics, the lognormal mean and standard deviation had to be calculated from the available mean and standard deviation. This is not ideal as these conversions are only correct if the random variable is truly lognormally distributed, which based on previous observations (recall the statistical test performed in the section Resonance Clustering) may not be true. Furthermore, amplitude statistics were not tracked during the traditional processing, preventing comparisons between traditional and AutoHVSR processing in terms of amplitude and its variability. Nonetheless, the comparisons are the best that can be made by the authors. Figure \ref{fig:9}a demonstrates excellent agreement between the exponentiated lognormal mean frequencies identified by AutoHVSR and traditional processing. This excellent agreement is quantified with an $R^2$ of 0.994 and a RMSE of 0.049. Figure \ref{fig:9}b demonstrates only moderate agreement between the lognormal standard deviation of frequency identified by AutoHVSR and traditional processing. The authors believe the disagreement in terms of the lognormal standard deviation is most-likely the result of compounding differences between the two processing styles, particularly across analysts who were making subjective decisions about which individual windows to reject, rather than an indictment of the AutoHVSR algorithms performance more generally. Despite the challenges of the Canterbury, New Zealand dataset aforementioned the AutoHVSR algorithm demonstrates strong performance.

\begin{figure}[t!]
    \centering
	\includegraphics[width=1.0\textwidth]{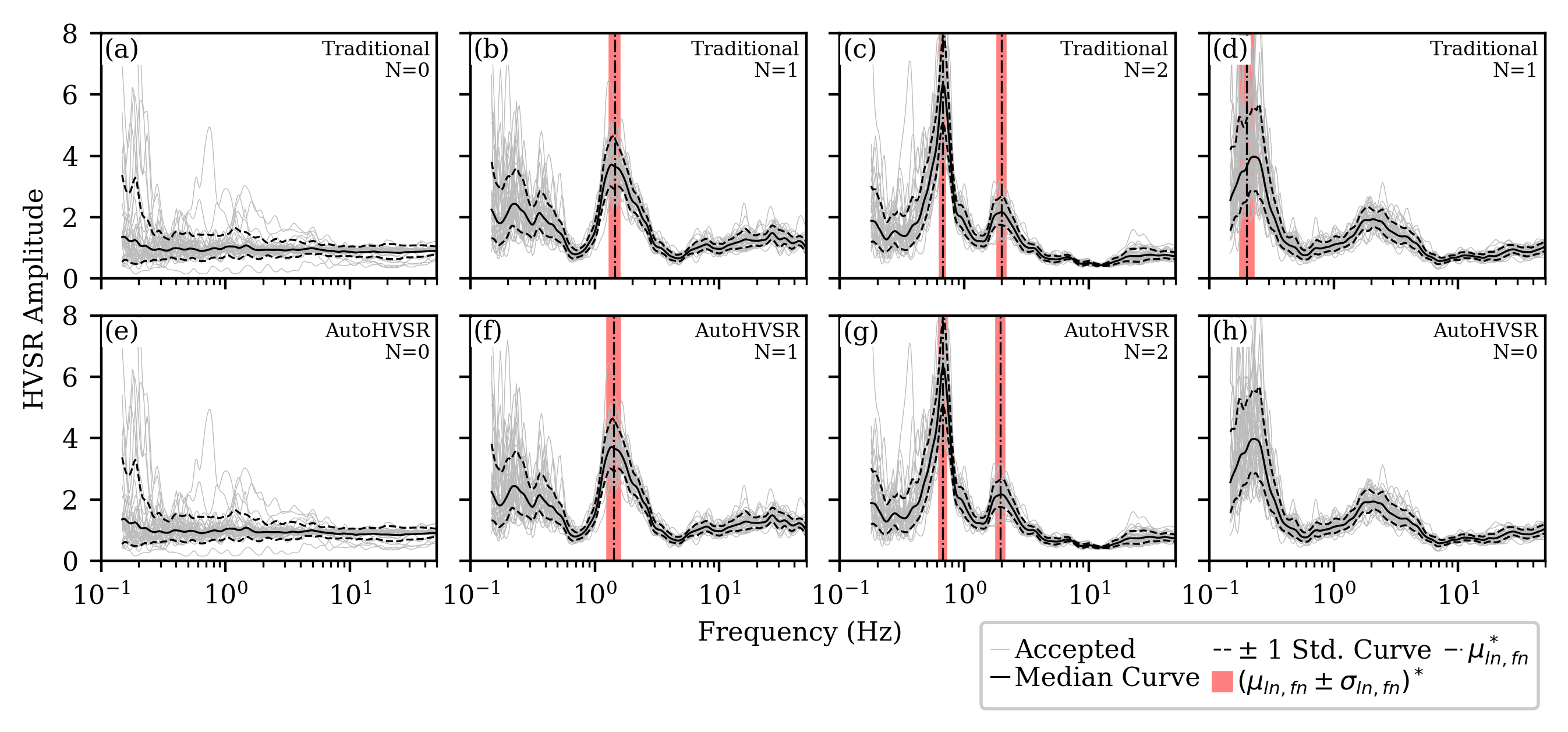}
	\caption{Examples from the Canterbury dataset prepared using traditional processing (top row) and the AutoHVSR algorithm (bottom row). The four examples were selected to represent the four main categories represented in the dataset's confusion matrix (Figure \ref{fig:8}) and includes cases where the number of resonances determined from AutoHVSR agree panels (a) through (c) \& (e) through (g) and disagree (d \& h). Each panel includes HVSR curves from the accepted time windows (thin dark line), lognormal median HVSR curve (dark solid line), and $\pm$ one lognormal standard deviation HVSR curve (dark dashed lines). Note that for these examples no HVSR curves were rejected. The distribution of each resonance in terms of frequency is also shown in the form of the exponential lognormal mean frequency ($\mu_{ln,fn}^*$) (dash dot dark line) and one lognormal standard deviation range ($(\mu_{ln,fn}\pm\sigma_{ln,fn})^*$). Note that each resonance's distribution in terms of amplitude is not shown because this information was not recorded during the traditional processing of the Canterbury, New Zealand dataset.}
	\label{fig:10}
\end{figure}

To present a more visual comparison, Figure \ref{fig:10} illustrates four examples from the Canterbury, New Zealand dataset representative of HVSR measurements from the four main categories of the dataset's confusion matrix (Figure \ref{fig:8}) These include cases where the number of resonances identified by AutoHVSR processing agree panels (a) through (c) \& (e) through (g) and where they do not (d \& h). Similar to Figures \ref{fig:6} and \ref{fig:7}, the top row of Figure \ref{fig:10} illustrates the results from traditional processing and the bottom row the results of processing with AutoHVSR. Shown in each panel are the HVSR curves from the accepted time windows (thin dark line), lognormal median HVSR curve (solid dark line), and $\pm$ one lognormal standard deviation HVSR curves (dashed dark lines). Note that for these examples we have not rejected any HVSR curves, however it is likely that during traditional processing some HVSR curves were rejected interactively in \textit{geopsy} to determine the resonance statistics presented. The distribution of each resonance in terms of frequency is also shown in the form of the exponential lognormal mean frequency ($\mu_{ln,fn}^*$) (dash dot dark line) and one lognormal standard deviation range ($(\mu_{ln,fn}\pm\sigma_{ln,fn})^*$). Note that each resonance's distribution in terms of amplitude is not shown because this information was not recorded during the traditional processing of the Canterbury, New Zealand dataset. As expected, good agreement is observed for the cases where the same number of resonances have been identified using both AutoHVSR and traditional processing [i.e., panels (a \& e), (b \& f), and (c \& g)]. For the final set of panels (i.e., d \& h) the traditional processing identified a HVSR resonances at approximately 0.2 Hz, however AutoHVSR did not. There are two considerations as to why AutoHVSR did not detect this resonance, first is that it is located at the low-frequency processing limit thereby making only about half of the resonance observable and second the peak does not pass the SESAME clarity criteria as it only satisfies 4 of the 6 clarity criteria. Both of these considerations make it difficult for AutoHVSR to confidently classify the resonance at 0.2 Hz and therefore it has not. Undoubtedly, different analysts will disagree about whether the peak in Figure \ref{fig:10}d is clear or not, which raises a very important point in regard to the future use of AutoHVSR.

As we have shown on both the 1109-member dataset used to develop AutoHVSR and on the 129-member dataset from Canterbury, New Zealand, AutoHVSR predicts the number of resonances and the statistics associated with those resonances with high accuracy. However, AutoHVSR is not 100\% accurate, this in part is due to the limitations of the methodology but more fundamentally, due to the fact that the classification of HVSR resonances is a subjective process about which different experienced analysts may disagree. Therefore, it is imperative that as AutoHVSR is utilized in the future to quickly process large datasets that those involved carefully review the results and use site-specific information (i.e., local geological constraints and prior HVSR measurements) when possible to confirm the results obtained. This requirement may appear to some to limit the key advantages of automated HVSR processing, that is, if each result from AutoHVSR needs to be reviewed carefully why not process using the traditional approach from the beginning. To respond to this potential criticism, consider the following hypothetical situation. We have 300 HVSR measurements from a city microzonation study. Processing these individually will take approximately 2 minutes per HVSR measurements or 600 minutes (10 hours) to process all 300 measurements. On a modern laptop using only a single thread AutoHVSR will be able to process this data in under 4 minutes (this includes HVSR calculation time). The analyst must then review the results to determine which of the HVSR measurements must be reprocessed by hand. From timings performed as part of this study, this will take a well-trained analyst under 7 seconds per HVSR measurement or 35 minutes to complete the review of all 300 measurements. Using the complex Canterbury, New Zealand dataset as a ``worst-case'' scenario approximately 13\% of the 300 or 39 HVSR measurements will need to be processed manually, which using 2 minutes per HVSR measurement, will take approximately 78 minutes. Therefore, to completely process the 300 HVSR measurements using the traditional method will take 10 hours to complete as compared to 2 hours (4 minutes for initial processing + 35 minutes for review + 78 minutes for manual reprocessing) at the worst using AutoHVSR (speed up of 5). This example clearly demonstrates that using AutoHVSR with a detailed review and as-needed reprocessing is substantially faster than traditional processing alone. In addition to speed, it also important to note that AutoHVSR processing is completely reproducible, consistent between analysts and across large datasets, and will continue to be improved as more HVSR data becomes available for building larger and more-diverse training sets.

\section*{Extensions to Azimuthally-Rotated and Earthquake HVSR}

\begin{figure}[t!]
    \centering
	\includegraphics[width=0.5\textwidth]{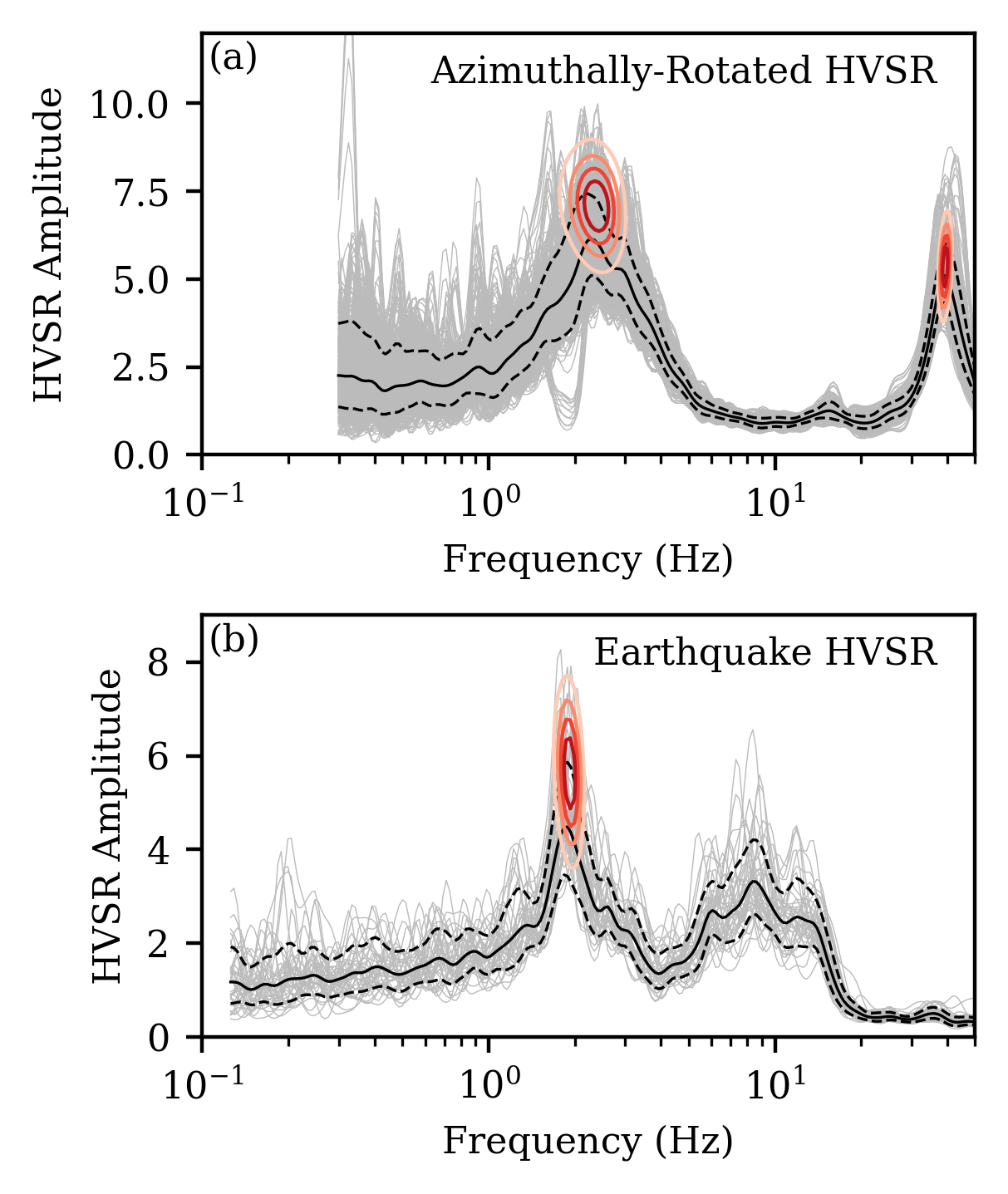}
	\caption{Application of the AutoHVSR algorithm to data dissimilar to that which it was trained including (a) mHVSR where the horizontal components are azimuthally-rotated and (b) HVSR computed from earthquake recordings (i.e., eHVSR). Measurements are from different sites and any similarity is coincidental. Each panel includes HVSR curves from the accepted time windows (thin dark line), lognormal median HVSR curve (solid dark line), and $\pm$ one lognormal standard deviation HVSR curve (dark dashed lines). The distribution of each significant resonance's frequency and amplitude are also shown.}
	\label{fig:11}
\end{figure}

The AutoHVSR algorithm has been developed using HVSR measurements derived from 3-component microtremor measurements where the horizontal components have been combined using the geometric mean. This section demonstrates that the AutoHVSR algorithm has sufficient generality to perform well on HVSR measurements produced using different data preparation procedures and sources. In particular, Figure \ref{fig:11}a shows an HVSR measurement derived from microtremor data where the horizontal components have been azimuthally rotated, and Figure \ref{fig:11}b show an HVSR measurement derived from earthquake recordings (eHVSR). The 3-component microtremor data used to produce Figure \ref{fig:11}a was selected from the AutoHVSR testing set. The earthquake motions in Figure \ref{fig:11}b are those selected by Tao and Rathje (\citeyear{tao_insights_2019}) and used in other small-strain seismic site response studies at the Garner Valley Downhole Array site \citep{teague_measured_2018, vantassel_multi-reference-depth_2019}. Importantly, the measurements shown in Figure \ref{fig:11}a and \ref{fig:11}b are from different sites and any similarity is coincidental. Shown in each panel of Figure \ref{fig:11} is the HVSR curves from the accepted time windows (thin dark line), lognormal median HVSR curve (thick dark line), and $\pm$ one lognormal standard deviation range (thick dashed lines). The distribution of each significant resonance's frequency and amplitude are also shown. Figure \ref{fig:11} demonstrates that the AutoHVSR algorithm, while developed using microtremor data where the horizontal components have been combined using the geometric mean, can be extended to microtremor data where the horizontal components are azimuthally rotated and to HVSR measurements of earthquake recordings. The authors anticipate that the ability of AutoHVSR to processing mHVSR and eHVSR quickly and consistently will be of significant interest to those in the earthquake engineering community incorporating HVSR into the ground motion prediction equations (GMPEs) and other estimates of site-specific seismic hazard.

\section*{Conclusions}

This work proposes the AutoHVSR algorithm for fully-automating the processing of HVSR measurements including those with zero, one, or multiple clear resonances. The AutoHVSR algorithm accepts microtremor or earthquake recordings and user-defined HVSR processing parameters as input and returns the HVSR calculations, statistics on the HVSR curve, and statistics on each automatically identified HVSR resonant frequency. The AutoHVSR algorithm demonstrates excellent performance by correctly determining the number of HVSR resonances for over 99\% of the 1109-member dataset used for its development and testing and by predicting the mean resonant frequency of those correctly identified resonances with a RMSE of 0.05 Hz. Furthermore the AutoHVSR algorithm was able to produce these predictions in 13 minutes (including HVSR calculation time) compared to the 30 hours required for traditional processing (a speed up of 138). The robustness of the AutoHVSR algorithm was assessed further using a previously analyzed dataset from Canterbury, New Zealand. The Canterbury, New Zealand dataset is a challenging test case as it was prepared by a group of analysts with the assistance of local geology and with information about resonances identified using other HVSR measurements in the local vicinity (i.e., information that is not available to AutoHVSR). Yet, the AutoHVSR algorithm demonstrates strong performance on the Canterbury, New Zealand dataset, correctly determining the number of resonances for 113 of the 129 measurements (\textgreater87\%) and predicting the mean resonant frequency of the correctly identified resonances with a RMSE of 0.05. Finally, while the AutoHVSR algorithm was developed using only microtremor measurements where the horizontal components were combined using the geometric mean, it is shown to extend without modification to microtremor HVSR where the two horizontal components are rotated azimuthally and to HVSR measurements from earthquake recordings. The authors hope that by developing and making this algorithm publicly available we can increase the consistency and reproducibility of HVSR processing across analysts, enable future studies of the physics underlying HVSR with multiple resonances, and facilitate the scientific community's ability to analyze large-scale datasets with the HVSR technique.

\subsection*{Acknowledgements}

Hyperparameter tuning of the extreme-gradient boosted trees model was performed on the Texas Advanced Computing Center (TACC) resource Frontera using resources provided through the DesignSafe-CyberInfrastructure \citep{rathje_designsafe_2017} and an allocation provided to the first author. The AutoHVSR algorithm has been made publicly available in v0.3.0 of HVSRweb \citep{vantassel_hvsrweb_2021}, it is hosted on resources provided by DesignSafe-CI \citep{rathje_designsafe_2017} and can be accessed at \href{https://hvsrweb.designsafe-ci.org/}{https://hvsrweb.designsafe-ci.org/}. The figures in this paper were created using Matplotlib 3.5.2 \citep{hunter_matplotlib_2007} and Inkscape 1.1.2.

\subsection*{Conflict of Interest}

The authors declare no competing interests.

\bibliographystyle{plainnat}
\bibliography{autohvsr}

\end{document}